# Fourier-Legendre expansion of the one-electron density-matrix of ground-state two-electron atoms

Sébastien RAGOT* and María Belén RUIZ[#]

*Laboratoire Structure, Propriété et Modélisation des Solides (CNRS, Unité Mixte de Recherche 85-80). École Centrale Paris, Grande Voie des Vignes, 92295 Chatenay-Malabry, France

[#]Deparment of Theoretical Chemistry, Friedrich-Alexander University Erlangen-Nürnberg, Egerlandstraße 3, 91058 Erlangen, Germany

## Abstract

The density-matrix $\rho(\mathbf{r}, \mathbf{r}')$ of a spherically symmetric system can be expanded as a Fourier-Legendre series of Legendre polynomials $P_l(\cos\theta = \mathbf{r}.\mathbf{r}'/rr')$. Application is here made to harmonically trapped electron pairs (*i.e.* Moshinsky's and Hooke's atoms), for which exact wavefunctions are known, and to the helium atom, using a near-exact wavefunction. In the present approach, generic closed form expressions are derived for the series coefficients of $\rho(\mathbf{r}, \mathbf{r}')$. The series expansions are shown to converge rapidly in each case, with respect to both the electron number and the kinetic energy. In practice, a two-term expansion accounts for most of the correlation effects, so that the correlated density-matrices of the atoms at issue are essentially a linear functions of $P_l(\cos\theta) = \cos\theta$. For example, in the case of the Hooke's atom: a two-term expansion takes in 99.9 % of the electrons and 99.6 % of the kinetic energy. The correlated density-matrices obtained are finally compared to their determinantal counterparts, using a simplified representation of the density-matrix $\rho(\mathbf{r}, \mathbf{r}')$, suggested by the Legendre expansion. Interestingly, two-particle correlation is shown to impact the angular delocalization of each electron, in the one-particle space spanned by the $\mathbf{r}$ and $\mathbf{r}'$ variables.

* Author to whom correspondence should be addressed. Electronic mail: at_home@club-internet.fr

# I. Introduction

The one-electron (or first-order) reduced density-matrix is a rich and fascinating quantum-mechanical function, with connections to various experimental results. Unfortunately, as soon as electron correlation is involved, the one-electron reduced density-matrix cannot anymore be formulated in closed form and this, even for the simplest two-electron systems and even when an exact wavefunction is available.

In this respect, the helium atom is the historical home of electron correlation studies. Though the very exact ground-state wavefunction is not known, extraordinarily accurate approximations have been developed [1]. Aside this "true" atom, several model systems have been invented, which in contrast allow for obtaining an exact wavefunction. Perhaps the most famed is the Hooke's atom, having two electrons harmonically trapped about a nucleus, which repel one another via a Coulomb term. In other words, the Hamiltonian (in a.u.) is

$$H = -\tfrac{1}{2}\nabla_1^2 - \tfrac{1}{2}\nabla_2^2 + \tfrac{1}{2}k(r_1^2 + r_2^2) + 1/r_{12}. \qquad (1)$$

This model was introduced years ago by Kestner and Sinanoglu [2], looking for a tractable model of correlated electrons for which an exact solution can be formulated, at least in power series. A ground-state wavefunction they formulated is $\psi = \mathcal{N} e^{-\frac{k^{1/2}}{2}(r_1^2 + r_2^2)} f(r_{12})$, wherein $f$ is shown to be *almost* linear in $r_{12}$. Later, Kais *et al.* have realized that the exact wave function could be written in closed form, in the specific case of $k = 1/4$ [3], whereby $f$ is *exactly* linear in $r_{12}$. This specific model ($k = 1/4$), or 'hookium' [4], has therefore attracted attention, offering a springboard to a number of exact properties for an harmonically trapped electron pair [4, 5, 6,7,8,9]. Since then, the eigenenergies and eigenfunctions of the Hooke's atom have been extensively investigated, beyond the specific case of $k = 1/4$ [10].

In addition to the Hooke's atom, other model atoms have been proposed [6, 7, 9, 11]. In some cases, the Coulomb term in eq. (1) is replaced by a quadratic potential or by an inverse square law potential [6, 7, 9]. As another example, a system of two electrons bound by a quadratic potential, with an additional linear term in the inter-electronic potential, was investigated by Ghosh and Samanta. [11]. Incidentally, while such systems are often contemplated as model atoms, they can more generally be considered as a paradigm of various physical systems, including quantum dots.

Referring back to hookium ($k = 1/4$): besides the wavefunction in position space, a number of properties can be obtained in closed form. Amongst them are the electron pair and charge densities, and kinetic energy tensors (see a summary in [5]). Interestingly, even the momentum-space wavefunction can be obtained in closed form, which in turn allows for deriving analytical forms of position and momentum intracules [4]. Moreover, the correlation energy of hookium, $E_c$ = 38.44 m$E_h$ ($E_{exact}$ = 2 $E_h$, see Ref. [4]), is close enough to the usual correlation energy of two-electron systems to make it an interesting basis for comparison with a real two-electron atom.

Unfortunately, the exact one-electron (spinless) density-matrix $\rho(\mathbf{r};\mathbf{r}')$ can manifestly not be obtained in closed form [5], in spite of interesting attempts to simplify its structure [6,9]. As a consequence, properties related to the off-diagonal part of the density-matrix (e.g. momentum properties) cannot be easily obtained, an exception being the reciprocal form factor [12]. An ad-hoc expansion of the momentum density has yet been proposed, which at least allows for a practical calculation [13]. However, said expansion converges very slowly and is not very instructive *per se*.

More generally, $\rho(\mathbf{r};\mathbf{r}')$ cannot be obtained in closed form for systems of electrons interacting via a Coulomb potential, even for the smallest atoms. At the best, the exact density-matrix can be represented in integral form or, more classically, expanded in a basis of orbital products. Therefore, there is a need for a convenient representation of the exact one-electron density-matrix, specifically for two-electron systems, for which exact or accurate wavefunctions is available.

To this aim, it is proposed to expand the exact density-matrix as a series of Legendre polynomials (section III), as in a previous work of Cioslowski and Buchowiecki [14]. Yet, application is here made to two sorts of harmonically trapped pairs of fermions as well as to helium atom, and analytical closed form expressions of the series coefficients of the density-matrix are derived in each case. Next, the convergence of the expansions is studied (section IV) and the converged density-matrices are finally compared to uncorrelated approximations (section V), using a remarkably simple, yet interesting representation of the density-matrix. Helpful definitions are first recalled.

## II. Fourier-Legendre expansion of the "spinless" density-matrix

### A. One-electron functions

First, the one-electron reduced density-matrix $\gamma(x;x')$ is usually defined with respect to a $N$-electron wave function $\psi$ [15, 16] as:

$$\gamma(x;x') = N \int \psi(x,x_2,...,x_N) \psi^*(x',x_2,...,x_N) dx_2...dx_N, \qquad (2)$$

where $x_i$ variables include both space and spin variables. Integrating eq. (2) over spin variables leads to the "spinless" density-matrix, which function is central to this paper:

$$\rho(\mathbf{r},\mathbf{r}') = \int [\gamma(x;x')]_{s=s'} ds. \qquad (3)$$

Hence, the resulting function (hereafter the "density-matrix", for short) describes an average spinless particle in the one-particle space.

Closely related, the electron charge density reduces to the diagonal $\rho(\mathbf{r}) = \rho(\mathbf{r},\mathbf{r})$, wherein information related to $\mathbf{u} = \mathbf{r} - \mathbf{r}'$ is discarded, at variance with off-diagonal properties, such as the momentum density $n(\mathbf{p})$ and the kinetic energy [15], which are determined by information related to **r** ≠ **r'**, and wherein the local information (**r** = **r'**) participates only in an average manner.

Finally, the one-electron density-matrix is subjected to the normalization condition

$$\int \rho(\mathbf{r},\mathbf{r})d\mathbf{r} = \int \rho(\mathbf{r})d\mathbf{r} = N .\qquad(4)$$

**B. Fourier-Legendre series expansion**

Usually, the density-matrix can be decomposed as a sum of orbital products:

$$\rho(\mathbf{r},\mathbf{r}') = \sum_i n_i \varphi_i(\mathbf{r})\varphi_i(\mathbf{r}'),\qquad(5)$$

where $n_i$ stands for the occupation number of orbitals $\varphi_i$. While a determinantal density-matrix can be written as a finite sum of orbital products, e.g. only one for $^1S$ state of a two-electron atom, a correlated approach may involve an infinity of products, reflecting that **r** and **r'** are in fact not separable.

A partial separation can yet be achieved using a Fourier-Legendre expansion, as in Ref. [14]. For example, the density-matrix $\rho$(**r**, **r'**) of a *spherically symmetric* system can be expanded as a series of Legendre polynomials $P_l(\cos\theta = \mathbf{r}.\mathbf{r}'/rr')$, such that series coefficients solely depend on $r$ and $r'$. In other words, the charge density $\rho(\mathbf{r})$ of spherical systems is independent from the orientation of **r**, so that $\rho(\mathbf{r}) \equiv \rho(r)$, but $\rho(\mathbf{r},\mathbf{r}')$ depends on $r$, $r'$ and the angle $\theta$ between **r** and **r'**. This applies in general to the symmetric part of the charge density operator of atoms, see e.g. eq. (4-66) of ref. [15] and, in particular, to the $^1S$ states of two-electron atoms, as exemplified below.

Now, since Legendre polynomials form a complete orthogonal system over the interval [-1, 1], any function $F(x)$ may in principle be expanded in terms of Legendre polynomials as [17]:

$$F(x) = \sum_{l=0}^{\infty} \alpha_l P_l(x)\qquad(6)$$

wherein

$$\alpha_l = \frac{2l+1}{2} \int_{-1}^{1} P_l(x) F(x) dx. \tag{7}$$

Making use of $x = \cos\theta = \mathbf{r}.\mathbf{r}'/rr'$, each coefficient in the series solely depends on $r$ and $r'$. Thus, the density-matrix can be expanded as

$$\rho(\mathbf{r},\mathbf{r}') = \sum_{l=0}^{\infty} \rho_l(r,r') P_l(\cos\theta). \tag{8}$$

Finally, with the above normalization choice, the charge density $\rho(\mathbf{r}) = \rho(\mathbf{r},\mathbf{r})$ reduces to

$$\rho(\mathbf{r}) = \rho(\mathbf{r},\mathbf{r}) = \sum_{l=0}^{\infty} \rho_l(r,r). \tag{9}$$

## III. Expansion of the Density-matrix in the (r, r') representation

### A. Moshinsky's atom

This point is first illustrated through a simple example, the Moshinsky's atom. Here, two fermions are bounded from a harmonic potential but repel via a quadratic potential $-\frac{1}{2}\lambda r_{12}^2 = -\frac{1}{2}\lambda(\mathbf{r}_2 - \mathbf{r}_1)^2$. This model is exactly solvable [18] and allows for obtaining all relevant distributions in closed form [15]. This makes it possible to compare the exact and determinantal density-matrices, respectively

$$\rho(\mathbf{r},\mathbf{r}') = 2\mathcal{N}^2 e^{-\frac{1}{4}(\alpha+\beta)(r^2+r'^2) - \frac{1}{2}(\alpha-\beta)\mathbf{r}.\mathbf{r}'}, \tag{10}$$

and

$$\rho_D(\mathbf{r},\mathbf{r}') \equiv \rho_D(r,r') = 2\mathcal{N}_D^2 e^{-\gamma(r^2+r'^2)/2}, \tag{11}$$

wherein $\alpha$ and $\beta$ and $\gamma$ are constants [13]. The determinantal matrix reduces to a single orbital product $\rho_D \propto e^{-\gamma r^2/2} e^{-\gamma r'^2/2}$, contrary to the exact density-matrix. Though the latter can be obtained in closed form, eq. (10), on may expand it as a series of Legendre polynomials, should it be for comparison with the other systems under consideration. In this respect, the exact matrix may be expanded by identifying the coupling term $e^{-\frac{1}{2}(\alpha-\beta)\mathbf{r}\cdot\mathbf{r}'} = e^{-\frac{1}{2}(\alpha-\beta)r r'\cos\theta}$ in eq. (10) to $F(x)$ in eq. (6), with $x = \cos\theta$. Therefore, one is able to recover the decomposition of eq. (8), with

$$\rho_l(r,r') = 2\mathcal{N}^2 (2l+1) e^{-\frac{1}{4}(\alpha+\beta)(r^2+r'^2)} (i)^l j_l\left(\tfrac{i}{2}(\alpha-\beta) r\, r'\right), \tag{12}$$

where $j_l(x)$ denotes a spherical Bessel function. In a variant, making use of modified Bessel functions $I_a(x) = i^{-a} J_a(ix)$, where $J_l(x)$ is for ordinary Bessel functions, eq. (12) rewrites

$$\rho_l(r,r') = 2\mathcal{N}^2 (2l+1) e^{-\frac{1}{4}(\alpha+\beta)(r^2+r'^2)} (-1)^l \sqrt{\frac{\pi}{(\alpha-\beta)r\,r'}} I_{l+1/2}\left(\tfrac{1}{2}(\alpha-\beta)r.r'\right). \tag{13}$$

Each coefficient of the series is thus manifestly real.

As shall be seen in section IV, the Legendre expansion converges very rapidly in this case and is essentially a linear function of $P_1(\cos\theta) = \cos\theta$ (for given $r$ and $r'$). In contrast, a determinantal approximation, see eq. (11), does clearly not depend on $\theta$. In other words, a Legendre expansion thereof reduces to a zero-order term.

A similar expansion can be contemplated for $^1S$ state of any two-electron atoms [16], as illustrated in the next sections, in reference to the hookium and helium.

**B. Hookium**

Since electrons are still bounded from the harmonic potential $\tfrac{1}{2}k(r_1^2+r_2^2)$, the wavefunction remains primarily Gaussian in nature. Although the Coulomb repulsion complicates the evaluation of the wavefunction [2], using $k = 1/4$ enables to write the wave function and charge density in closed form [3,5], that is

$$\psi(R_{12}, r_{12}) = \mathcal{N}\, e^{-R_{12}^2/2} e^{-r_{12}^2/8} (1 + r_{12}/2), \text{ and} \tag{14}$$

$$\rho(r) = \mathcal{N}^2 \frac{\pi\sqrt{2\pi}}{r} e^{-r^2/2}\left(7r + r^3 + \left(\frac{8}{\sqrt{2\pi}}\right) r e^{-r^2/2} + 4(1+r^2)\mathrm{erf}\left(\frac{r}{\sqrt{2}}\right)\right),$$

wherein $\mathcal{N} = \left(2\pi^{5/4}\left(5\pi^{1/2}+8\right)^{1/2}\right)^{-1}$, and $R_{12}$ and $r_{12}$ correspond to the centre-of-mass $\mathbf{R}_{12} = (\mathbf{r}_1 + \mathbf{r}_2)/2$, and the relative coordinate $\mathbf{r}_{12} = \mathbf{r}_2 - \mathbf{r}_1$, respectively.

From eqs. (2) and (3), we get an integral expression

$$\rho(\mathbf{r}_1, \mathbf{r}_1') = 2\mathcal{N}^2 \, e^{-(r_1^2 + r_1'^2)/4} \int e^{-r_2^2/2} (1 + r_{12}/2)(1 + r_{1'2}/2) \, d\mathbf{r}_2, \tag{15}$$

which involves odd powers of the interelectronic distances $r_{12} = |\mathbf{r}_2 - \mathbf{r}_1|$ and $r_{1'2} = |\mathbf{r}_2 - \mathbf{r}_1'|$. Thus, four integrals have to be calculated, as noted in ref. [14]:

$$I_0 = \int e^{-r_2^2/2} \, d\mathbf{r}_2, \tag{16}$$
$$I_1(r_1) = \int e^{-r_2^2/2} (r_{12}) \, d\mathbf{r}_2,$$
$$I_{1'}(r_{1'}) = I_1(r_{1'}), \text{ and}$$
$$I_2(r_1, r_{1'}, \cos\theta_{11'}) = \int e^{-r_2^2/2} \tfrac{1}{4}(r_{12} r_{1'2}) \, d\mathbf{r}_2.$$

Next, integrals $I_0$, $I_1$ and $I_{1'}$ are easily evaluated (e.g. by changing the integration variable). We obtain accordingly:

$$I_0 = (2\pi)^{3/2}, \text{ and} \tag{17}$$
$$I_1(r_1) = I_{1'}(r_1) = \pi\left(2e^{-r_1^2/2} + \frac{(2\pi)^{1/2}(1 + r_1^2)\mathrm{erf}(r_1/\sqrt{2})}{r_1}\right).$$

However, the last integral $I_2$ cannot be evaluated in closed form due to the product $|\mathbf{r}_2 - \mathbf{r}_1||\mathbf{r}_2 - \mathbf{r}_1'|$. To arrive at a tractable form thereof, one may expand the interelectronic distance as

$$r_{12} = \sum_{l=0}^{\infty} f_l(s_{12}, g_{12}) P_l(\cos\theta_{12}), \tag{18}$$

where $f_l$ is a function of the radial distances $s_{12}$ and $g_{12}$, respectively the smaller and the greater of $r_1$ and $r_2$ (using the same definitions as in ref. [19]). It can be verified that the functions $f_q$ write explicitly [19, 20]:

$$f_l(s_{12}, g_{12}) = \left(\frac{1}{2l+3}\right)\frac{s_{12}^{l+2}}{g_{12}^{l+1}} - \left(\frac{1}{2l-1}\right)\frac{s_{12}^{l}}{g_{12}^{l-1}}. \tag{19}$$

Next, using the addition theorem for spherical harmonics, we obtain from eq. (18):

$$r_{12} = \sum_{l=0}^{\infty} \frac{4\pi}{2l+1} f_l(s_{12}, g_{12}) \sum_{m=-l}^{l} Y_l^m(\theta_1, \varphi_1) Y_l^{m*}(\theta_2, \varphi_2). \tag{20}$$

Thus, integral $I_2$ rewrites as

$$I_2(r_1, r_{1'}, \cos\theta_{11'}) = \sum_{l,l'} \sum_{m,m'} \left(\frac{4\pi}{2l+1}\right)\left(\frac{4\pi}{2l'+1}\right) \times \tag{21}$$
$$\int e^{-r_2^2/2} \tfrac{1}{4}\left\{f_l(s_{12}, g_{12}) f_{l'}(s_{1'2}, g_{1'2}) Y_l^m(1) Y_l^{m*}(2) Y_{l'}^{m'}(2) Y_{l'}^{m'*}(1')\right\} d\mathbf{r}_2,$$

wherein $Y_l^m(\theta_1, \varphi_1) \equiv Y_l^m(1)$. Decomposing now $d\mathbf{r}_2$ as $r_2^2 dr_2 d\Omega_2 = r_2^2 dr_2 \sin\theta_2 d\theta_2 d\varphi_2$ and integrating over $\Omega_2$, one takes advantage of orthogonal spherical harmonics. The residual sum can be reduced using once more the addition theorem, leading to

$$I_2(r_1, r_{1'}, \cos\theta_{11'}) = \sum_l \left(\frac{4\pi}{2l+1}\right) P_l(\cos\theta_{11'}) \int \tfrac{1}{4} r_2^2 dr_2 e^{-r_2^2/2} f_l(s_{12}, g_{12}) f_l(s_{1'2}, g_{1'2}). \tag{22}$$

Accordingly, $I_2$ can finally be decomposed as $I_2(r_1, r_{1'}, \cos\theta_{11'}) = \sum_l I_{2,l}(r_1, r_{1'}) P_l(\cos\theta_{11'})$, with coefficients $I_{2,l}$ defined by

$$I_{2,l}(r_1, r_{1'}) = \left(\frac{4\pi}{2l+1}\right) \int \tfrac{1}{4} r_2^2 dr_2 e^{-r_2^2/2} f_l(s_{12}, g_{12}) f_l(s_{1'2}, g_{1'2}). \tag{23}$$

Thus, the density-matrix as defined in eq. (15) finally decomposes as in eq. (8), *i.e.* $\rho(\mathbf{r}_1, \mathbf{r}_1') = \sum_l \rho_l(r_1, r_{1'}) P_l(\cos\theta_{11'})$, wherein the zero-order term ($l = 0$) is given by

$$\rho_0(r_1, r_{1'}) = 2\mathcal{N}^2 \, e^{-(r_1^2 + r_{1'}^2)/4} \{I_0 + I_1(r_1) + I_1(r_{1'}) + I_{2,0}(r_1, r_{1'})\}, \tag{24}$$

and the subsequent terms ($l > 0$) by

$$\rho_{l>0}(r_1, r_{1'}) = 2\mathcal{N}^2 \, e^{-(r_1^2 + r_{1'}^2)/4} I_{2,l>0}(r_1, r_{1'}). \tag{25}$$

Finally, to complete the density-matrix expansion, closed form expressions have to be obtained for the integrals $I_{2,l}$. In this regard, the definition of $s_{12}$ and $g_{12}$ (the smaller and the greater of $r_1$ and $r_2$) makes $I_{2,l}(r_1, r_{1'})$, eq. (23), to decompose as the sum of three integrals. To illustrate this, we further introduce variables $s$ and $g$, denoting the smaller and the greater of $r_1$ and $r_{1'}$ (or $r$ and $r'$ for conciseness). Accordingly, while $s_{12}$ and $g_{12}$ discriminates amongst radial distances involved in the two-particle space, $s$ and $g$ relate to the one-particle space spanned by $r \equiv r_1$ and $r' \equiv r_{1'}$. Having this in mind, $I_{2,l}(r,r')$ rewrite explicitly as

$$I_{2,l}(r,r') = \left(\frac{4\pi}{2l+1}\right) \left\{ \int_0^s \frac{1}{4} r_2^2 dr_2 \ e^{-r_2^2/2} f_l(r_2,s) f_l(r_2,g) \right. \tag{26}$$

$$+ \int_s^g \frac{1}{4} r_2^2 dr_2 \ e^{-r_2^2/2} f_l(s,r_2) f_l(r_2,g)$$

$$\left. + \int_g^\infty \frac{1}{4} r_2^2 dr_2 \ e^{-r_2^2/2} f_l(s,r_2) f_l(g,r_2) \right\}$$

Finally, a closed form expression can be obtained for $I_{2,l}(r,r')$, as to be discussed now. In fact, each of the three integrals in eq. (26) can be evaluated in terms of incomplete gamma functions $\Gamma(a, x)$ [21]. In particular, we may use the formula:

$$\int_{r_1}^\infty r_2^n e^{-\beta r_2^2} dr_2 = \frac{1}{2} \frac{\Gamma(\frac{n+1}{2}, \beta r_1^2)}{\beta^{(n+1)/2}}, \qquad n \in N \tag{27}$$

which is valid for $n$ positive, even and odd. For negative powers (even or odd), we may instead make use of the formula:

$$\int_{r_1}^\infty \frac{1}{r_2^n} e^{-\beta r_2^2} dr_2 = \frac{1}{2} \beta^{n/2} \Gamma(-\frac{n}{2}, \beta r_1^2), \qquad n \in N \tag{28}$$

Next, taking into account the following relations

$$\int_0^s dr_2 = \int_0^\infty dr_2 - \int_s^\infty dr_2 \tag{29}$$

$$\int_s^g dr_2 = \int_s^\infty dr_2 - \int_g^\infty dr_2$$

and keeping in mind that the incomplete gamma function $\Gamma(a, x)$ generalizes the complete gamma function ($\Gamma(a, 0) = \Gamma(a)$), we arrive, for each of the three integrals of eq. (26), at a combination of both incomplete and complete Gamma functions. Put together, we finally obtain for the coefficients $I_{2,l}$

$$\begin{aligned}
I_{2,l}(r_1, r_{1'}) = \frac{\pi}{2(2l+1)} & \left\{ \frac{1}{(2l+3)^2} \frac{1}{\beta^{l+7/2}} \left( \Gamma\left(\frac{7}{2}+l\right) - \Gamma\left(\frac{7}{2}+l, \beta s^2\right) \right) \frac{1}{g^{l+1} s^{l+1}} \right. \\
& - \frac{1}{(2l+3)(2l-1)} \frac{1}{\beta^{l+5/2}} \left( \Gamma\left(\frac{5}{2}+l\right) - \Gamma\left(\frac{5}{2}+l, \beta s^2\right) \right) \left( \frac{1}{g^{l-1} s^{l+1}} + \frac{1}{g^{l+1} s^{l-1}} \right) \\
& + \frac{1}{(2l-1)^2} \frac{1}{\beta^{l+3/2}} \left( \Gamma\left(\frac{3}{2}+l\right) - \Gamma\left(\frac{3}{2}+l, \beta s^2\right) \right) \frac{1}{g^{l-1} s^{l-1}} \\
& + \frac{1}{(2l+3)^2} \frac{1}{\beta^2} \left( \Gamma(2, \beta s^2) - \Gamma(2, \beta g^2) \right) \frac{s^{l+2}}{g^{l+1}} \\
& - \frac{1}{(2l+3)(2l-1)} \frac{1}{\beta} \left( \Gamma(1, \beta s^2) - \Gamma(1, \beta g^2) \right) \frac{s^{l+2}}{g^{l-1}} \\
& - \frac{1}{(2l+3)(2l-1)} \frac{1}{\beta^3} \left( \Gamma(3, \beta s^2) - \Gamma(3, \beta g^2) \right) \frac{s^l}{g^{l+1}} \\
& + \frac{1}{(2l-1)^2} \frac{1}{\beta^2} \left( \Gamma(2, \beta s^2) - \Gamma(2, \beta g^2) \right) \frac{s^l}{g^{l-1}} \\
& + \frac{\beta^{l-1/2}}{(2l+3)^2} \Gamma\left(\frac{1}{2}-l, \beta g^2\right) g^{l+2} s^{l+2} \\
& - \frac{\beta^{l-3/2}}{(2l+3)(2l-1)} \Gamma\left(\frac{3}{2}-l, \beta g^2\right) (g^l s^{l+2} + g^{l+2} s^l) \\
& \left. + \frac{\beta^{l-5/2}}{(2l-1)^2} \Gamma\left(\frac{5}{2}-l, \beta g^2\right) g^l s^l \right\}
\end{aligned} \tag{30}$$

wherein $\beta = 1/2$, and $s$ and $g$ denote the smaller and the greater of $r_1$ and $r_{1'}$, respectively. Accordingly, eqs. (15 - 17) and (30) define the coefficients of the series expansion of the density-matrix in closed form. Though the obtained expression, notably eq. (30), is somewhat long, it does not involve more than incomplete Gamma functions and is easily computed. In a variant, integrals $I_{2,l}(r, r')$ can else be easily evaluated for particular values of $l$.

For the sake of illustration, the zero-order term $\rho_0(r, r')$ obtained is depicted in FIG. 1A, following the same convention as in ref. [22]. The expected spherical pattern is obtained: its maximum is located at $r = r' = 0$. Subsequent terms of the expansion are drawn in FIGS. 1B - C. In contrast with $\rho_0(r, r')$, the first-order term $\rho_1(r, r')$ and the second-order term $\rho_2(r, r')$ have a maximum shifted from the origin. Besides the differences in magnitude, $\rho_2(r,r')$ is slightly shifted and contracted, compared to $\rho_1(r,r')$. Accordingly, the terms $\rho_{l>0}(r,r')$ which bear the angular dependence between vectors **r** and **r'** are somehow more localized in space. The values of the contours further denote a fast convergence, as to be discussed in section IV.

<Note: Please insert FIGS. 1A - C here>

## C. Helium atom

In the case of helium, the interacting electrons are now placed in a central Coulomb potential. This situation may nicely be reflected in the historical Hylleraas wavefunction [23]:

$$\psi(r_1, r_2, r_{12}) = \mathcal{N} \, e^{-\zeta(r_1+r_2)}(1+\alpha r_{12}). \tag{31}$$

Minimizing the corresponding energy, as in ref. [4], leads to -2.891 120 7 a.u., which is approximately 0.0126 a.u. too high but nevertheless renders about 99.6 % of the exact energy. Parameters obtained are $\alpha = 0.365\,796\,2$ and $\zeta = 1.849\,684\,5$, that is, not too far from their ideal cusp values of 1/2 and 2, respectively. In comparison, the coefficient obtained within the mean-field approach, *i.e.* $\psi_0(r_1, r_2) = \mathcal{N}_0 \, e^{-\zeta(r_1+r_2)}$, is $\zeta = 2 - 5/16 = 1.687\,5$.

Beside its simplicity, the Hylleraas wave function leads to a meaningful comparison with hookium, since the wavefunction is in both cases a linear function of $r_{12}$. Thus, the same method as for hookium can be applied. More generally, the same method can be applied to any system for which the (trial) wavefunction writes as

$$\psi(r_1, r_2, r_{12}) = \mathcal{N} \, \varphi(r_1)\varphi(r_2)(1+\alpha r_{12}). \tag{32}$$

In the present case, the density-matrix writes formally:

$$\rho(\mathbf{r}_1, \mathbf{r}_1') = 2\mathcal{N}^2 \, \varphi(r_1)\varphi(r_{1'})^* \int |\varphi(r_2)|^2 (1+\alpha r_{12}+\alpha r_{1'2}+\alpha^2 r_{12} r_{1'2}) \, d\mathbf{r}_2 \tag{33}$$
$$= 2\mathcal{N}^2 \, e^{-\zeta(r_1+r_{1'})} \int e^{-2\zeta r_2}(1+\alpha r_{12}+\alpha r_{1'2}+\alpha^2 r_{12} r_{1'2}) \, d\mathbf{r}_2.$$

Again, four integrals are needed:

$$I_0 = \int e^{-2\zeta r_2}(1)\, d\mathbf{r}_2, \tag{34}$$

$$I_1(r_1) = \alpha \int e^{-2\zeta r_2}(r_{12})\, d\mathbf{r}_2,$$

$$I_{1'}(r_{1'}) = I_1(r_1'), \text{ and}$$

$$I_2(r_1, r_{1'}, \cos\theta_{11'}) = \alpha^2 \int e^{-2\zeta r_2}(r_{12} r_{1'2})\, d\mathbf{r}_2.$$

The three integrals $I_0$, $I_1$, and $I_{1'}$ are easily evaluated, using for instance the expansion of $r_{12}$ discussed earlier. Next, just as in hookium, integral $I_2$ can *in fine* be decomposed as

$$I_2(r_1, r_{1'}, \cos\theta_{11'}) = \sum_l I_{2,l}(r_1, r_{1'}) P_l(\cos\theta_{11'}), \tag{35}$$

wherein

$$I_{2,l}(r_1, r_{1'}) = \left(\frac{4\pi}{2l+1}\right) \int r_2^2 dr_2 e^{-2\zeta r_2} f_l(s_{12}, g_{12}) f_l(s_{1'2}, g_{1'2}). \tag{36}$$

Accordingly, the expansion of eq. (8) is recovered, wherein the zero-order term is given by

$$\rho_0(r_1, r_{1'}) = 2\mathcal{N}^2\, \varphi(r_1)\varphi(r_{1'})^* \{I_0 + I_1(r_1) + I_1(r_{1'}) + I_{2,0}(r_1, r_{1'})\} \tag{37}$$
$$= 2\mathcal{N}^2\, e^{-\zeta(r_1+r_{1'})} \{I_0 + I_1(r_1) + I_1(r_{1'}) + I_{2,0}(r_1, r_{1'})\}$$

and the subsequent terms by

$$\rho_{l>0}(r_1, r_{1'}) = 2\mathcal{N}^2\, \varphi(r_1)\varphi(r_{1'})^* I_{2,l>0}(r_1, r_{1'}) \tag{38}$$
$$= 2\mathcal{N}^2\, e^{-\zeta(r_1+r_{1'})} I_{2,l>0}(r_1, r_{1'}).$$

Next, a general closed form expression of $I_{2,l}(r, r')$ can be obtained, here again. In the present case, integrals $I_0$ and $I_1$ are given by

$$I_0 = \int e^{-2\zeta r_2} d\mathbf{r}_2 = \frac{\pi}{\zeta^3} \tag{39}$$

and

$$I_1(r_1) = \pi\alpha \left\{ \frac{1 - \zeta^2 r_1^2}{\zeta^5 r_1} - e^{-2\zeta r_1}\left(\frac{2 + 4\zeta^3 r_1^3 + 7\zeta r_1 + 8\zeta^2 r_1^2}{2\zeta^5 r_1}\right) \right\} \tag{40}$$

Then, coefficients $I_{2,l}$ decompose once more as a sum of three integrals and each of the three integrals can be evaluated using the incomplete Gamma function, namely,

$$\int_s^\infty r_2^n e^{-\beta r} dr = \frac{1}{\beta^{n+1}} \Gamma(n+1, \beta s). \quad n \in \mathbb{Z} \tag{41}$$

which is valid for a positive or negative integer $n$.

Following just the same method as in the case of hookium, we finally arrive at

$$\begin{aligned} I_{2,l}(r_1, r_{1'}) = \frac{\pi\alpha^2}{2l+1} &\Bigg\{ \frac{1}{(2l+3)^2} \frac{1}{\beta^{2l+7}} \left(\Gamma(2l+7) - \Gamma(2l+7, \beta s)\right) \frac{1}{g^{l+1} s^{l+1}} \\ &- \frac{1}{(2l+3)(2l-1)} \frac{1}{\beta^{2l+5}} \left(\Gamma(2l+5) - \Gamma(2l+5, \beta s)\right) \left( \frac{1}{g^{l-1} s^{l+1}} + \frac{1}{g^{l+1} s^{l-1}} \right) \\ &+ \frac{1}{(2l-1)^2} \frac{1}{\beta^{2l+3}} \left(\Gamma(2l+3) - \Gamma(2l+3, \beta s)\right) \frac{1}{g^{l-1} s^{l-1}} \\ &+ \frac{1}{(2l+3)^2} \frac{1}{\beta^4} \left(\Gamma(4, \beta s) - \Gamma(4, \beta g)\right) \frac{s^{l+2}}{g^{l+1}} \\ &- \frac{1}{(2l+3)(2l-1)} \frac{1}{\beta^2} \left(\Gamma(2, \beta s) - \Gamma(2, \beta g)\right) \frac{s^{l+2}}{g^{l-1}} \\ &- \frac{1}{(2l+3)(2l-1)} \frac{1}{\beta^6} \left(\Gamma(6, \beta s) - \Gamma(6, \beta g)\right) \frac{s^l}{g^{l+1}} \\ &+ \frac{1}{(2l-1)^2} \frac{1}{\beta^4} \left(\Gamma(4, \beta s) - \Gamma(4, \beta g)\right) \frac{s^l}{g^{l-1}} \\ &+ \frac{\beta^{2l-1}}{(2l+3)^2} \Gamma(1-2l, \beta g) \, g^{l+2} s^{l+2} \\ &- \frac{\beta^{2l-3}}{(2l+3)(2l-1)} \Gamma(3-2l, \beta g) \left(g^l s^{l+2} + g^{l+2} s^l\right) \\ &+ \frac{\beta^{2l-5}}{(2l-1)^2} \Gamma(5-2l, \beta g) \, g^l s^l \Bigg\} \end{aligned} \tag{42}$$

wherein $\beta = 2\zeta$ and $s$ and $g$ denote the smaller and the greater of $r_1$ and $r_{1'}$, respectively. Eqs. (33 - 34) and (37 - 40) define the coefficients of the series expansion of the density-matrix.

Accordingly, the coefficients of the series expansion of the density-matrix are determined in closed form for each of the two-electron systems at issue.

## IV. Convergence

To illustrate the convergence of the above expansion, radial densities $4\pi r^2 \rho_l(r,r)$ corresponding to $l$-order terms ($l = 0$ to 3) are depicted in FIG. 1A-C, for each of the model atoms at issue. Contributions of the coefficients of the density-matrix expansion to both the electron number and the kinetic energies are aggregated in Table 1.

## A. Moshinsky's atom

First, concerning the Moshinsky's atom, we choose $k = 1/4$ and $\lambda = 0.241296 \times 1/4$ for a practical application, whereby $k > 2\lambda$, as required for the state to remain stable. Such values make the Moshinsky's atom resemble hookium inasmuch as the same average radius of the fermion cloud is obtained, i.e. $\langle r \rangle = 1.7445$ a.u. Notwithstanding the differences of magnitude in FIG. 2 (see the scale factors reported in FIG. 2), the successive contributions $4\pi r^2 \rho_l(r,r)$ are essentially similar though shifted along the $r$-axis. We note the gap between two consecutive maxima.

As to numerical values (see Table 1): the contributions ($l = 0$ to $3$) to either the electron number or the exact kinetic energy $T = \frac{3}{4}\left(\sqrt{k} + \sqrt{(k-2l)}\right)$ converge rapidly. Yet, contributions corresponding to $l > 1$ are not significant vis-à-vis the order of magnitude of kinetic energy correlation ($10^{-2}$ a.u.). Altogether, such contributions amounts to about 0.05% of the kinetic energy (about 3 $10^{-4}$ a.u.). The contributions to the kinetic energies have been evaluated via the formula:

$$T_l = -\tfrac{1}{2}\int \Delta_\mathbf{r} \rho_l(r,r') P_l(\cos\theta)\big|_{\mathbf{r}=\mathbf{r'}} d\mathbf{r} . \tag{43}$$

Two conclusions raise. First, the Fourier-Legendre expansion converges very rapidly. Second, for given $r$ and $r'$, the density-matrix is essentially a linear function of $P_1(\cos\theta) = \cos\theta$.

## B. Hookium

Second, concerning hookium, each of the contributions $4\pi r^2 \rho_l(r,r)$ drawn in FIG. 3 is conveniently rescaled such that the curves are contiguous at $r \approx 2$ a.u. The scale factors used (see the legend in FIG. 3) illustrate the differences of orders of magnitude of the consecutive terms. Besides, we note the gap between the first and second maxima, which markedly differs from the subsequent gaps, in contrast with the previous example. Maxima are nevertheless located not far from $\langle r \rangle = 1.7445$ a.u. The global trend confirm the description of FIG. 1A - C, that is, $\rho_{l+1}(r,r)$ is slightly shifted and contracted, compared with $\rho_l(r,r)$.

**< Please insert FIG. 3 >**

Next, as to be seen in table 1, the contributions depicted in FIG. 3 nicely converge, at least as long as first momentum moments ($N$ and $T$) are involved. In particular, it can be verified that the electron numbers $N_l$ (for l = 0 - 3) correspond exactly to the values given in [14]. Yet, the convergence is not as rapid as in the previous case, due to the coulomb repulsion. As to the corresponding kinetic energies, integration was carried out analytically for $l = 0$ and $l = 1$, and numerically for higher order terms, taking care of numerical instabilities occurring at small $r$. The following percentages of $T_{exact} = 0.664\ 418$ a.u. are successively recovered: 93.93% with $T_0$, 99.60% with $T_0 + T_1$, 99.90 with $T_0 + T_1 + T_2$, and 99.96% with $T_0 + T_1 + T_2 + T_3$).

In comparison, a Hartree-Fock or a Kohn-Sham approximation to the one-electron matrix [13, 24] leads to about 95 - 96 % of the exact kinetic energy only. Thus, most of the correlation kinetic energy (and more generally most of the correlation effects) is reflected in a two-term expansion (*i.e.* 99.60% of $T_{exact}$). Besides, this conclusion can manifestly be somewhat generalized, owing to the results given in Ref. [14], in the case of $k = 1/100$ (corresponding to $\omega = 1/10$ in Ref. [14]), as to the electron numbers $N_l$ (for l = 0 - 4), namely, 1.82935, 0.16958, 0.00093, 0.00011, and 0.00002.

## C. Helium atom

Successive contributions $4\pi r^2 \rho_l(r,r)$ are depicted in FIG. 4, for $l = 0, 1, 2$ and 3. Said contributions are rescaled so that the curves are contiguous at about 1.3 a.u. Again, we point at the gap between the first and second maxima, compared with the subsequent gaps. Maxima are located near $\langle r \rangle = 0.8968$ a.u. The scale factors used here (1, 97, 3 900, and 32 000) reveal an even better convergence than in the previous case. In fact, the very first term of the expansion ($l = 0$) already provides a much more reliable approximation of the converged momentum moments (see table 1), namely 99.30% of $N$ (= 2) and 98.62% of $T$ (= 2.891 121, as from the trial wavefunction of eq. (31)). Thus, the angular dependence of the one-matrix is less pregnant in helium compared with harmonic case. This will be further discussed later.

**< Please insert FIG. 4 >**

Next, adding one term in the expansion of the density-matrix leads to $T_0 + T_1$, which covers 99.91% of $T$. A four-term expansion yields 99.99% thereof ($T_0 + T_1 + T_2 + T_3$).

For the sake of comparison, the simple mean-filed approximation $\varphi_0(r) \propto e^{-\zeta r}$ leads to about 98.5% of the converged value ($T = 2.891\ 121$). Similarly, the limit HF or the exact KS matrix leads to about 99 % of the *exact* kinetic energy [13]. Accordingly, a two-term expansion of the density-matrix takes in most of the correlation kinetic energy (*i.e.* about 1% of the total kinetic energy), here too.

To conclude this section, a Legendre expansion of the density-matrix in the (**r**, **r'**) representation is manifestly efficient, at least for the atoms considered here. Interestingly, most of electron correlation effects in the one-particle space is captured in a two-term expansion of the density-matrix, which is therefore essentially a linear function of $P_1(\cos\theta) = \cos\theta$.

## V. Comparison with uncorrelated matrices

In section III, we have derived explicit expressions for the coefficients $\rho_l(r, r')$ in the expansion of the density-matrix of two-electron atoms; the efficiency of the expansion scheme was discussed in section IV. Now, we shall put on view how the two-particle correlation in the wavefunction impacts the one-electron matrix.

To make this more intelligible, one may for instance integrate the radial terms in the expansions discussed above and focus on the residual dependence of the density-matrix on the angle $\theta$ between **r** and **r'**. Basically, setting $r = r'$ in the expansion of eq. (8), i.e. $\rho(\mathbf{r},\mathbf{r'}) = \sum_l \rho_l(r,r') P_l(\cos\theta)$, multiplying it by $4\pi r^2$ and integrating over $r$ leads to the following, partially integrated representation of the density-matrix:

$$\hat{\rho}(\theta) = \sum_l \left( \int 4\pi r^2 \rho_l(r,r) dr \right) P_l(\cos\theta) \qquad (44)$$
$$= \sum_l n_l P_l(\cos\theta),$$

the normalization condition implying $\hat{\rho}(0) = N = 2$ here. Thus, the "$\theta$" representation establishes a drastic simplification of the density-matrix, which accordingly reduces to a simple polynomial of $\cos\theta$.

In terms of numerical values, the density-matrices obtained for the model atoms at issue (denoted by *My*, *Ho*, and *He* below) are:

$$\hat{\rho}_{My}(\theta) = 1.95984\ P_0(\cos\theta) + 0.03971 P_1(\cos\theta) + 0.00045\ P_2(\cos\theta) + ... \qquad (45)$$
$$= 1.95995 + 0.03971\cos\theta + 0.00034\cos 2\theta + ...,$$

$\hat{\rho}_{Ho}(\theta) = 1.95141 + 0.04752\cos\theta + 0.00089\cos 2\theta + ...$, and

$\hat{\rho}_{He}(\theta) = 1.98611 + 0.01362\cos\theta + 0.00023\cos 2\theta + ...$.

The functions $\hat{\rho}_X(\theta)$ are next compared in FIG. 5: the simplicity of the said "$\theta$" representation makes it possible to appreciate critical resemblances and differences between the cases at stake. Here, all curves show a bell-shaped curve centered at $\theta = 0$, with a small amplitude (compare the values taken at $\theta = 0$ and at $\theta = \pm\pi$). In contrast, since KS or HF orbitals are spherically symmetric by definition, the corresponding density-matrix solely depends on $r$ and $r'$ and not on $\theta$. Accordingly, in the "$\theta$" representation, a determinantal density-matrix of a ground-state two-electron system is flat and reduces to $\hat{\rho}_D(\theta) = 2$.

A representation of the "correlation" density-matrix can thus be introduced:

$$\Delta\hat{\rho}(\theta) = \hat{\rho}(\theta) - \hat{\rho}_D(\theta) = \sum_l (n_l - 2\delta_{l,0}) P_l(\cos\theta), \tag{46}$$

which makes it explicit that electron correlation results in redistributing a (small) fraction of electrons from level $l = 0$ to higher angular levels $l > 0$. Note that in practice, $\Delta\hat{\rho}(\theta)$ is simply obtained by translating the exact matrix $\hat{\rho}(\theta)$, at least for the atoms considered here. FIG. 5 can therefore be seen as a representation of the correlation density-matrix too.

The patterns shown therein indicates that electron correlation manifestly results in slightly confining the density-matrix about $\theta = 0$.

**< Please insert FIG. 5 >**

This point can in fact be seen as a consequence of *angular* correlation of an electron pair, as to be illustrated now through a simple model. Consider for instance the simple trial wavefunction:

$$\psi(r_1, r_2, r_{12}) = \mathcal{N}\, \varphi(r_1)\varphi(r_2)(1 - \alpha \hat{\mathbf{r}}_1 \cdot \hat{\mathbf{r}}_2) = \mathcal{N}\, \varphi(r_1)\varphi(r_2)(1 - \alpha \cos\theta_{12}), \tag{47}$$

wherein $\alpha$ is a positive and real, but likely small. This wavefunction includes angular correlation in a naïve manner; yet, it takes larger values when $\theta_{12}$ tends to $\pm\pi$, in accordance with an angular correlation mechanism. Now, assuming that the orbital $\varphi$ is spherically symmetric and normalized, the corresponding one-matrix is

$$\rho(\mathbf{r}_1, \mathbf{r}_1') = 2\mathcal{N}^2\, \varphi(r_1)\varphi(r_{1'})^* (1 + \alpha^2 \hat{\mathbf{r}}_1 \cdot \hat{\mathbf{r}}_{1'}/3) \tag{48}$$
$$= 2\mathcal{N}^2\, \varphi(r_1)\varphi(r_{1'})^* \left(1 + \frac{\alpha^2}{3}\cos\theta_{11'}\right).$$

Therefore, the one-electron matrix above takes lower values for $\theta_{11'} \equiv \theta = \pi$ and has a maximum for $\theta_{11'} \equiv \theta = 0$, independently from $r$ and $r'$ (in agreement with the patterns in FIG. 5). We further note that, in the example above, the density-matrix is (exactly) as a linear function of $P_1(\cos\theta) = \cos\theta$.

Next, let us shortly refer back to eq. (5), $\rho(\mathbf{r},\mathbf{r}') = \sum_i n_i \varphi_i(\mathbf{r})\varphi_i(\mathbf{r}')$, wherein the density matrix writes a sum of products of amplitudes, considered at different points **r** and **r'**. Accordingly, it can be interpreted as a two-point probability amplitude, yet pertaining to one particle only. In other words, the density-matrix teaches an average electron being delocalized at two different locations **r** and **r'** and therefore includes information about a delocalized electron. Now, at correlated level, the two-point probability amplitude in question depends on $\theta_{11'}$ and has a maximum for $\theta_{11'} = 0$, at variance with a non-correlated description. Thus, the propensity of an electron to delocalize at different locations **r** and **r'** depends on the relative orientation thereof. In other words, the correlation of a pair of electrons impacts the behavior of each electron in the pair (and notably how said electron delocalizes). The most obvious manifestation of this is that the pair correlation impacts the kinetic energy of each electron.

## VI. Conclusions

To conclude this study, explicit Fourier-Legendre expansions of the density-matrix of two-electron atoms have been proposed, together with closed form expressions for the series coefficients. The series expansions have been shown to converge rapidly, at least as long as small momentum moments (e.g. $N$ and $T$) are involved. It appears that, in practice, a two-term expansion accounts for most of the correlation effects on the one-electron reduced density-matrix, at least for the two-electron atoms under consideration. Incidentally, such an expansion suggests a simplified representation of the density-matrix, *i.e.* the said "$\theta$" representation, wherein integration is carried out over the radial term, and only the angular variable survives. As to a practical application, the correlated density-matrices obtained were compared to their determinantal counterparts, by means of the said "$\theta$" representation of the density-matrix, which enables an original and simple illustration of correlation effects in the one-particle space. In particular, angular correlation of electrons is shown to slightly confine the angular delocalization of one electron *per se*.

## Acknowledgement

The present authors were not aware of Ref. [14]. They hereby thank the Referee for drawing its content to their attention during the referring process.

## Figure Captions

- FIG. 1A - C: Representation of $\rho_l(r,r')$ in the $(r, r')$ plane for hookium. FIG. 1A: $l = 0$; FIG. 1B: $l = 1$; and FIG. 1A: $l = 2$.

- FIG. 2: Moshinsky's atom: scaled representation of the contributions $4\pi r^2 \rho_l(r,r)$, for $l = 0$ to 3.

- FIG. 3: Hookium ($k = 1/4$): scaled representation of the contributions $4\pi r^2 \rho_l(r,r)$, for $l = 0$ to 3.

- FIG. 4: Helium: scaled representation of the contributions $4\pi r^2 \rho_l(r,r)$, for $l = 0$ to 3.

- FIG. 5: Comparison of density-matrices of two-electron atoms in the "$\theta$" representation. See text.

## Table Caption

Table 1: Convergence of electron number and kinetic energy from the Legendre expansion of the density-matrix - see text.

Figures

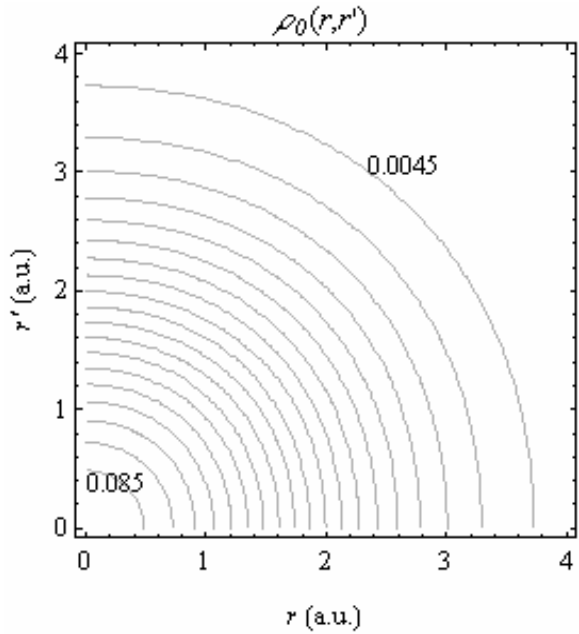

FIG. 1A

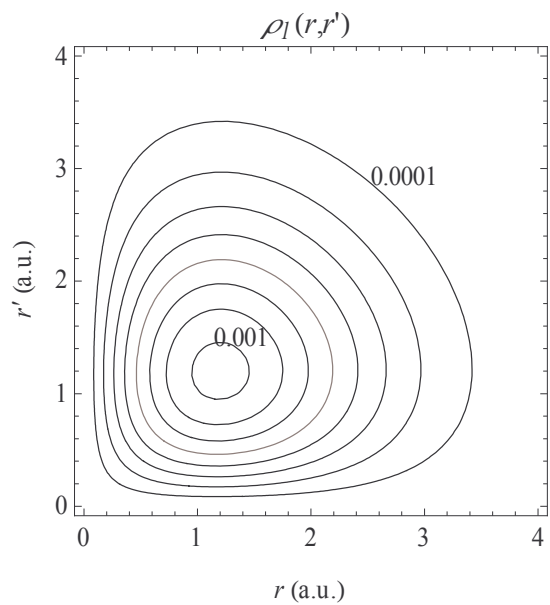

FIG. 1B

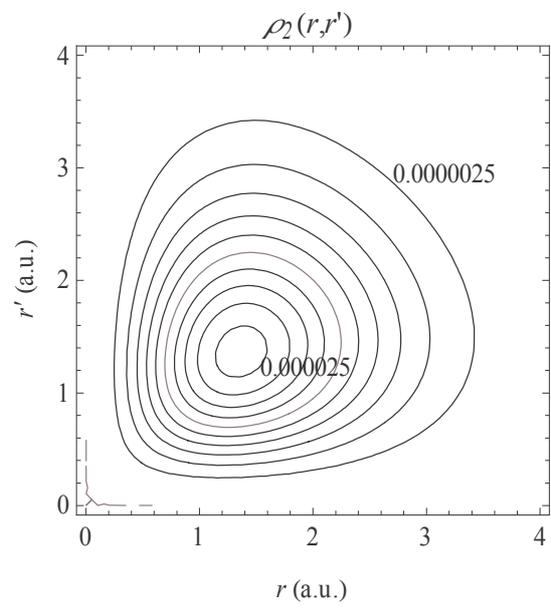

FIG. 1C

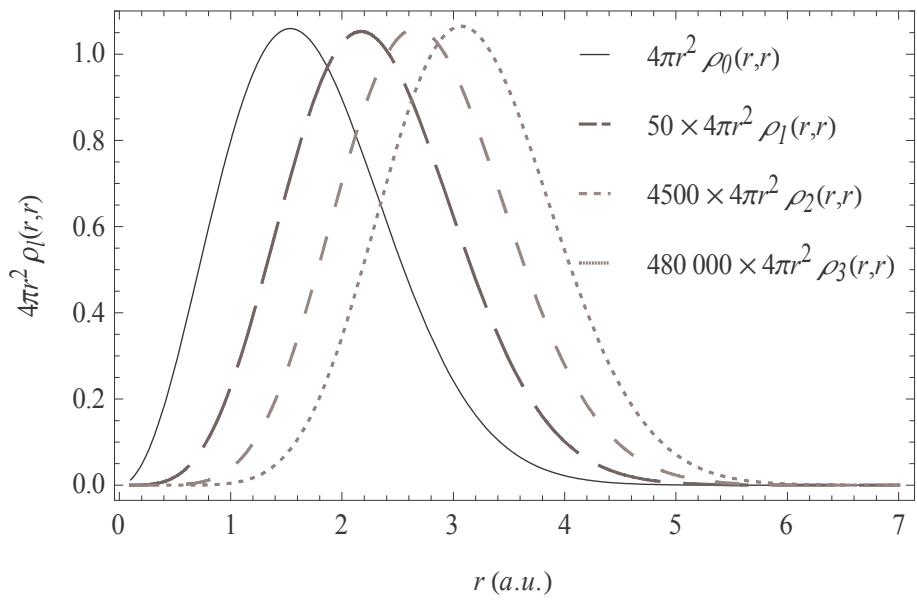

FIG. 2

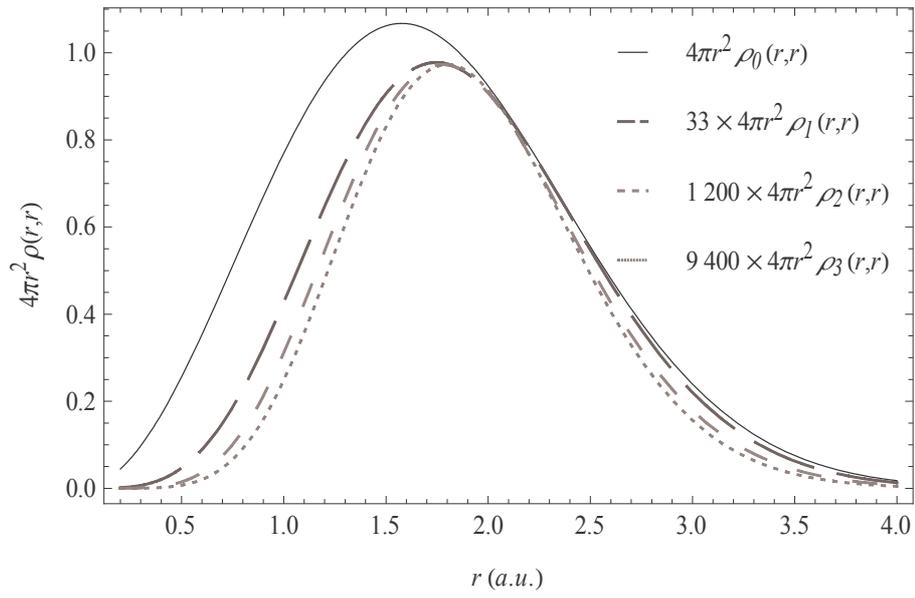

FIG. 3

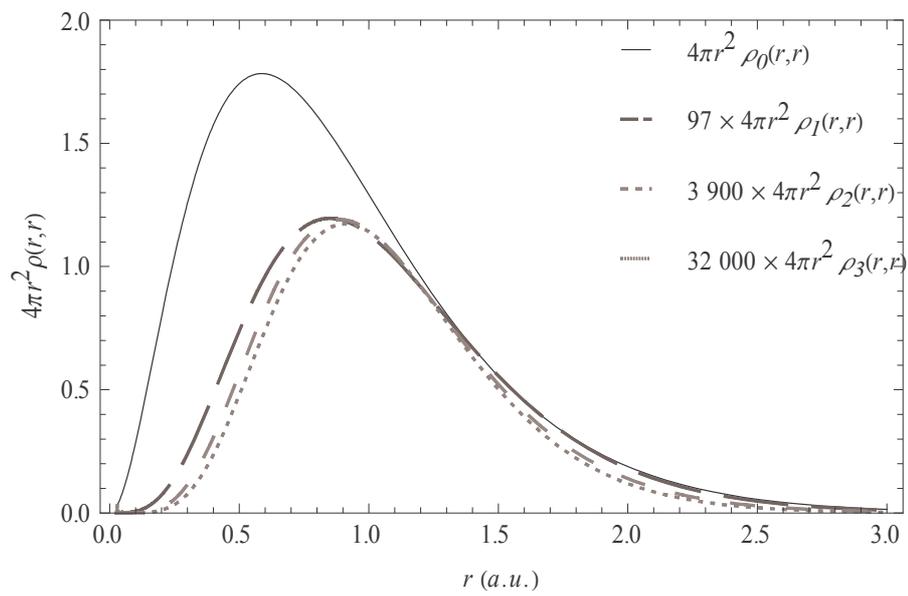

FIG. 4

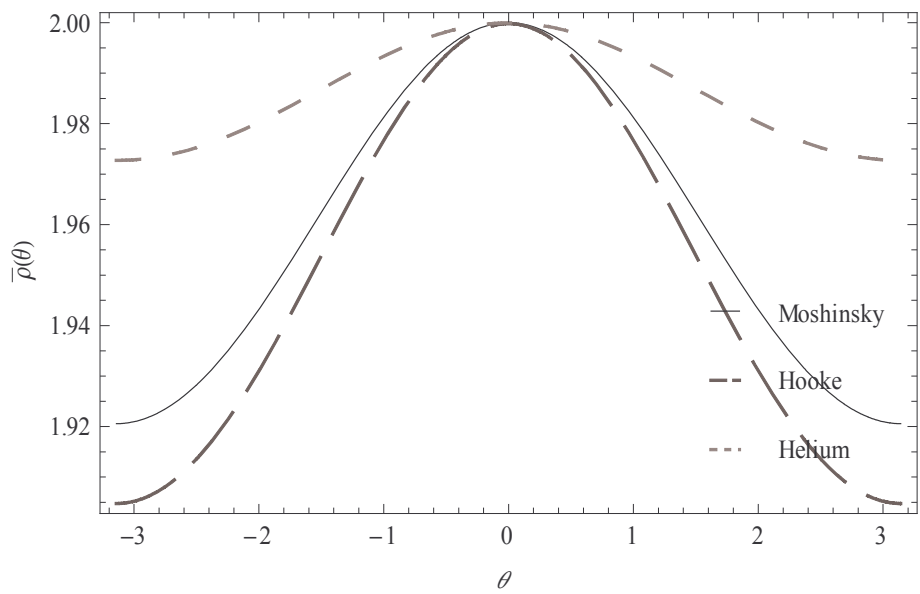

FIG. 5

Table

|  | Moshinsky | | Hookium | | Helium | |
|---|---|---|---|---|---|---|
|  | $\Sigma N_l$ | $\Sigma T_l$ | $\Sigma N_l$ | $\Sigma T_l$ | $\Sigma N_l$ | $\Sigma T_l$ |
| $l_{max} = 0$ | 1.95984 | 0.62336 | 1.95111 | 0.62408 | 1.98603 | 2.85130 |
| $l_{max} = 1$ | 1.99955 | 0.64441 | 1.99863 | 0.66178 | 1.99965 | 2.88867 |
| $l_{max} = 2$ | 2.00000 | 0.64474 | 1.99981 | 0.66373 | 1.99995 | 2.89050 |
| $l_{max} = 3$ | 2.00000 | 0.64474 | 1.99995 | 0.66414 | 1.99999 | 2.89087 |
| Converged Value | 2 | 0.64474 | 2 | 0.66442 | 2 | 2.89112 |